\documentclass{article}
 \usepackage{corrinchmargs}
 \usepackage{splat}
 \usepackage{hyperref}
 \usepackage{epic}
 \usepackage{eepic}

\makeatletter  

\newcommand{\comm}[1]{}

\def\uncompactdate#1#2#3#4#5#6#7#8{\ifcase#5#6\or January\or February\or March\or April\or May\or June\or July\or August\or September\or October\or November\or December\fi \space#7#8, #1#2#3#4}

\newcommand{\twodig}[1]{\ifcase#1 00\or01\or02\or03\or04\or05\or06\or07\or08\or09\else#1\fi}

\newcommand{\stretchset}[1]{\renewcommand{\baselinestretch}{#1}\@normalsize}

\renewcommand{\cleardoublepage}{\clearpage\if@twoside \ifodd\c@page\else
    \hbox{}\thispagestyle{plain}\newpage\if@twocolumn\hbox{}\newpage\fi\fi\fi}

\newcommand{\modestsections}{
    \let\part=\section
    \let\section=\subsection
        \renewcommand{\thesubsection}{\arabic{subsection}}
    \let\subsection=\subsubsection  
    \let\subsubsection=\paragraph
    \let\paragraph=\subparagraph
    \renewcommand{\subparagraph}[1]{\paragraph{##1}
        \typeout{You've used the subparagraph command, which is the same as
                 the paragraph command since you're using modest sections.}}
    \renewcommand\appendix{\par
        \setcounter{subsection}{0}%
        \setcounter{subsubsection}{0}%
        \renewcommand\thesubsection{\@Alph\c@subsection}}
}

\newcommand{\modestmaketitle}{
\def\@maketitle{\newpage
 \null
 \vskip 1.5em
 \begin{center}%
  {\Large \@title \par}%
  \vskip 1em
  {\large
   \lineskip .5em
   \begin{tabular}[t]{c}\@author
   \end{tabular}\par}%
  \vskip .7em
  {\large \@date}%
 \end{center}%
 \par
 \vskip 1em}
}

\newcommand{\arabicifpos}[1]{\@arabicifpos{\@nameuse{c@#1}}}
\newcommand{\@arabicifpos}[1]{\ifnum #1>0 \number #1\fi}

\def\rect{\@ifstar{\@blackrect}{\@rect}}

\def\square{\@ifstar{\@blacksquare}{\@square}}

\def\@square#1{\@rect(#1,#1)}

\def\@blacksquare#1{\@blackrect(#1,#1)}

\def\@rect(#1,#2){\makebox(0,0)%
{\begin{picture}(#1,#2)\put(0,0){\framebox(#1,#2){}}\end{picture}}}

\def\@blackrect(#1,#2){\makebox(0,0){\rule{#1\unitlength}{#2\unitlength}}}

\def\blackbox(#1,#2)#3{\makebox(#1,#2){\rule{#1\unitlength}{#2\unitlength}}}

\let\blackcircle=\@dot

\newsavebox{\object}
\newsavebox{\arrayputsrow}
\long\def\arrayputs(#1,#2)(#3,#4)#5#6#7{
    \savebox{\object}{#7}
    \savebox{\arrayputsrow}{\multiput(#1,0)(#3,0){#5}{\usebox{\object}}}
    \sbox{\object}{}
    \multiput(0,#2)(0,#4){#6}{\usebox{\arrayputsrow}}
    \sbox{\arrayputsrow}{}}

\def\horizonly@frameb@x#1{%
  \@tempdima\fboxrule
  \advance\@tempdima\fboxsep
  \advance\@tempdima\dp\@tempboxa
  \hbox{%
    \lower\@tempdima\hbox{%
      \vbox{%
        \hrule\@height\fboxrule
        \hbox{%
          #1%
          \vbox{%
            \vskip\fboxsep
            \box\@tempboxa
            \vskip\fboxsep}%
          #1}%
        \hrule\@height\fboxrule}%
                          }%
        }%
}

\newcommand{\horizonlyfboxes}{\let\@frameb@x=\horizonly@frameb@x}

{\begin{minipage}{#1}\begin{tabbing}}%
{\end{tabbing}\end{minipage}}

\newenvironment{tabpackage}
{\begin{minipage}{\columnwidth}\begin{tabbing}}%
{\end{tabbing}\end{minipage}}

\newcommand{\inputverb}[1]%
{\typeout{(#1 (verbatim)}\verbatiminput{#1}\typeout{)}}

\newcount\n
\def\repetition#1#2{\n=0\loop\ifnum\n<#2\advance\n by 1 #1\repeat}

\def\fewreps#1#2{\ifcase#2\or 
#1\or
#1#1\or
#1#1#1\or
#1#1#1#1\or
#1#1#1#1#1\or
#1#1#1#1#1#1\or
#1#1#1#1#1#1#1\or
#1#1#1#1#1#1#1#1\or
#1#1#1#1#1#1#1#1#1\or
#1#1#1#1#1#1#1#1#1#1\or
#1#1#1#1#1#1#1#1#1#1#1\or
#1#1#1#1#1#1#1#1#1#1#1#1\or
#1#1#1#1#1#1#1#1#1#1#1#1#1\or
#1#1#1#1#1#1#1#1#1#1#1#1#1#1\or
#1#1#1#1#1#1#1#1#1#1#1#1#1#1#1\or
#1#1#1#1#1#1#1#1#1#1#1#1#1#1#1#1\or
#1#1#1#1#1#1#1#1#1#1#1#1#1#1#1#1#1\or
#1#1#1#1#1#1#1#1#1#1#1#1#1#1#1#1#1#1\or
#1#1#1#1#1#1#1#1#1#1#1#1#1#1#1#1#1#1#1\or
#1#1#1#1#1#1#1#1#1#1#1#1#1#1#1#1#1#1#1#1\or
#1#1#1#1#1#1#1#1#1#1#1#1#1#1#1#1#1#1#1#1#1\or
#1#1#1#1#1#1#1#1#1#1#1#1#1#1#1#1#1#1#1#1#1#1\or
#1#1#1#1#1#1#1#1#1#1#1#1#1#1#1#1#1#1#1#1#1#1#1\or
#1#1#1#1#1#1#1#1#1#1#1#1#1#1#1#1#1#1#1#1#1#1#1#1\or
#1#1#1#1#1#1#1#1#1#1#1#1#1#1#1#1#1#1#1#1#1#1#1#1#1\or
#1#1#1#1#1#1#1#1#1#1#1#1#1#1#1#1#1#1#1#1#1#1#1#1#1#1\or
#1#1#1#1#1#1#1#1#1#1#1#1#1#1#1#1#1#1#1#1#1#1#1#1#1#1#1\or
#1#1#1#1#1#1#1#1#1#1#1#1#1#1#1#1#1#1#1#1#1#1#1#1#1#1#1#1\or
#1#1#1#1#1#1#1#1#1#1#1#1#1#1#1#1#1#1#1#1#1#1#1#1#1#1#1#1#1\or
#1#1#1#1#1#1#1#1#1#1#1#1#1#1#1#1#1#1#1#1#1#1#1#1#1#1#1#1#1#1\or
#1#1#1#1#1#1#1#1#1#1#1#1#1#1#1#1#1#1#1#1#1#1#1#1#1#1#1#1#1#1#1\or
#1#1#1#1#1#1#1#1#1#1#1#1#1#1#1#1#1#1#1#1#1#1#1#1#1#1#1#1#1#1#1#1\or
#1#1#1#1#1#1#1#1#1#1#1#1#1#1#1#1#1#1#1#1#1#1#1#1#1#1#1#1#1#1#1#1#1\or
#1#1#1#1#1#1#1#1#1#1#1#1#1#1#1#1#1#1#1#1#1#1#1#1#1#1#1#1#1#1#1#1#1#1\or
#1#1#1#1#1#1#1#1#1#1#1#1#1#1#1#1#1#1#1#1#1#1#1#1#1#1#1#1#1#1#1#1#1#1#1\or\else
\typeout{fewreps warning: Number of repetitions outside 0--35 range}\fi}

\makeatother

\newcounter{proglevel}
\newcounter{progline}
\newlength{\linenosep}
\newlength{\subprogindent}
\newlength{\progitemhangindent}
\newlength{\addprogitemsep}

\newenvironment{program}[1]%
{\setcounter{proglevel}{0}\setcounter{progline}{0}
 \settowidth{\linenosep}{\quad}
 \settowidth{\subprogindent}{\qquad}\settowidth{\progitemhangindent}{\quad}
 \setlength{\addprogitemsep}{0pt}
\newcommand{\setsubprogtabs}{#1\=\hspace*{\linenosep}
			     \hspace*{\value{proglevel}\subprogindent}\=
			     \hspace*{\progitemhangindent}\=\+\+\+\kill}
 \begin{tabpackage}\setlength{\tabbingsep}{0pt}\setsubprogtabs}%
{\end{tabpackage}}

\newenvironment{subprogram}%
{\stepcounter{proglevel}\pushtabs\<\<\<\-\-\-\setsubprogtabs}%
{\poptabs\addtocounter{proglevel}{-1}}

\newif\ifprogitemarg
\makeatletter
\def\progitem{\@ifnextchar [{\progitemargtrue\progitemwarg}%
{\progitemargfalse\progitemwarg[\proglinenum]}}
\makeatother

\let\prit=\progitem

\newlength{\progitemstrutheight}
\def\progitemwarg[#1]{\<\< #1 \'\>\rule{0pt}{\progitemstrutheight}%
\ifprogitemarg\else%
\addtocounter{progline}{-1}\refstepcounter{progline}\ignorespaces\fi}

\newcommand{\proglinenum}{\refstepcounter{progline}\theprogline}

\newcommand{\ifthenelse}[3]	
    {{\it if} #1 \begin{description}
				  \item[\hfill {\it then}] #2
		                  \item[\hfill {\it else}] #3
		              \end{description}}

\newcommand{\var}[1]{\mbox{\it #1}}

\newcommand{\Do}{{\bf do\ }}

\newcommand{\For}{{\bf for\ }}
\newcommand{\Endfor}{{\bf endfor\ }}
\newcommand{\To}{{\bf to\ }}

\newcommand{\If}{{\bf if\ }}
\newcommand{\Then}{{\bf then\ }}
\newcommand{\Else}{{\bf else\ }}

\newcommand{\NULL}{\mbox{\sc null}}

\makeatletter

\newcommand{\bibintro}[1]{%
\let\@origthebib=\thebibliography
\let\@origlist=\list
\def\thebibliography{\def\list{#1\par\bigskip\@origlist}\@origthebib}%
}

\makeatother

\newcommand{\inlineciteclosed}
  {\ifx\newblock\undefined\newcommand{\newblock}{\hskip .11em plus .33em minus .07em}\fi}

\inlineciteclosed

\newcounter{oscounter}

\newcommand{\ordinal}[1]{#1\ordsuffix{#1}}

\newcommand{\ordsuffix}[1] 
{\ifcase#1
th\or st\or nd\or rd\or th\or th\or th\or th\or th\or th\or
th\or th\or th\or th\else\setcounter{oscounter}{#1}\ordsuffixrecur\fi}

\newcommand{\ordsuffixrecur}
{\addtocounter{oscounter}{-10}\ifcase\value{oscounter}%
th\or st\or nd\or rd\or th\or th\or th\or th\or th\or th\else
\ordsuffixrecur\fi}

\newcommand{\soda}[1]
    {Proceedings of the \ordinal{#1} Annual {\ifnum#1>3ACM-\fi}SIAM
Symposium on Discrete Algorithms}

\newcommand{\spaa}[1]
    {Proceedings of the
     \ifcase#1\or 1989 \else{\ordinal{#1} Annual }\fi
     ACM Symposium on Parallel Algorithms and Architectures}

\newtheorem{theorem}{Theorem}

\newenvironment{proof}{\noindent {\it Proof.}\ \ }{\qed}
\makeatletter
\def\qed{
   {%
      \unskip
      \nobreak \hfil
      \penalty 50               
      \hskip 3em                
      \null \nobreak \hfil
      \qedbox
      \parfillskip=\z@skip
      \finalhyphendemerits=\z@  
      \endgraf                  
   }}
\makeatother
\newcommand{\qedbox}{\ \hfill\rule{1.5ex}{1.5ex}\vspace{\smallskipamount}}

\newcommand{\rigaddr}{Dept.\ of Computer Science\\
 Loyola University\\
 820 N. Michigan Ave.\\
 Chicago, IL\ \ 60611-2147}

\newcommand{\card}[1]{\left|#1\right|}

\newcommand{\istar}{i^{*}}
\newcommand{\jstar}{j^{*}}

\newcommand{\ihat}{\hat{i}}
\newcommand{\jhat}{\hat{j}}

\newcommand{\itilde}{\tilde{i}}
\newcommand{\jtilde}{\tilde{j}}

\newcommand{\head}{\var{head}}
\newcommand{\pretail}{\var{pretail}}
\newcommand{\tail}{\var{tail}}

\begin{document}

\date{} 

\title{Fast and Simple Computation \\ of All Longest Common Subsequences}

\author{Ronald I. Greenberg\\
 \rigaddr\\
 \url{http://www.cs.luc.edu/~rig}}

\maketitle

\begin{abstract}
 This paper shows that a simple algorithm produces the {\em
all-prefixes-LCSs-graph} in $O(mn)$ time for two input sequences of
size $m$ and $n$.  Given any prefix $p$ of the first input sequence
and any prefix $q$ of the second input sequence, all longest common
subsequences (LCSs) of $p$ and $q$ can be generated in time
proportional to the output size, once the all-prefixes-LCSs-graph has
been constructed.  The problem can be solved in the context of
generating all the distinct character strings that represent an LCS or
in the context of generating all ways of embedding an LCS in the two
input strings.

\end{abstract}

\begin{keyword}
 longest common subsequences, edit distance, shortest common supersequences
 \end{keyword}

\section{Background and Terminologies}
 \label{sec:intro}
 Let $A=a_1a_2\ldots a_m$ and $B=b_1b_2\ldots b_n$ with $m\leq n$ be
two sequences over an alphabet $\Sigma$.  Any sequence that can be
obtained by deleting some symbols of another sequence is referred to
as a {\em subsequence} of the original sequence.  A {\em common
subsequence} of $A$ and $B$ is a subsequence of both $A$ and $B$.  The
longest common subsequence (LCS) problem is to find a common
subsequence of greatest possible length.\footnote{ It is reasonable to
assume that the alphabet size $\card{\Sigma}$ is at most $m$, since
the actual value of symbols not present in the shorter string is
irrelevant.  Extraneous symbols can be culled efficiently if space
usage is not of concern, or hashing can be used to obtain a good
expected time with little space usage.}

\begin{figure}
 \centering
 \input{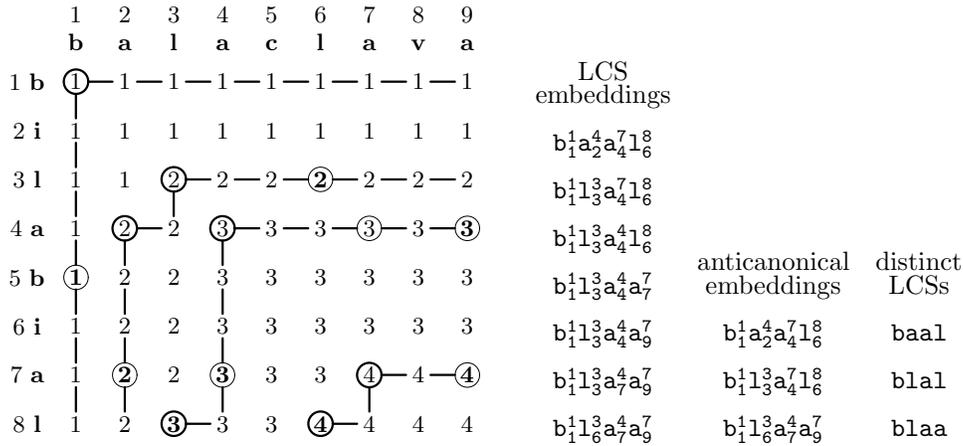}
 \caption{Listed are the seven different embeddings and three
anticanonical embeddings (corresponding to the three distinct LCSs)
for the strings $A={\tt bilabial}$ and $B={\tt balaclava}$.  (The
naive method of generating all LCSs for this pair of strings would
produce a list of length 100, because there would be many
duplications.)  In the matrix, the $[i,j]$ entry shows the rank
$L[i,j]$ as per~(\ref{eqn:naiverecur}).  The matches are
circled and are organized into contours as shown by the connecting
lines.  If a match is dominant, its circle is bold, and if the match
is antidominant, its rank is bold.  (Note that a match may be both
dominant and antidominant.)}
 \label{fig:example}
 \end{figure}

A pair of sequences may have many different LCSs.  In addition, a
single LCS may have many different {\em embeddings}, i.e., positions
in the two strings to which the characters of the LCS correspond.  We
may pick out a distinguished embedding for each distinct LCS, e.g.,
the {\em canonical embedding} has been defined to be the one in which
each character, starting from the beginning of the LCS, is assigned
matching positions in both sequences as small as
possible~\cite{Rick2000E}.  It is more convenient in this paper to
distinguish embeddings in which the matching positions are chosen as
large as possible (starting from the end of the LCS); let us call
these {\em anticanonical embeddings}.  Figure~\ref{fig:example} shows
an example pair of strings and the various LCS embeddings and
anticanonical embeddings.  (The matrix in the figure will be explained
later.)

A few other terminologies and notations that will be useful are as
follows.  We use $A_i$ to represent the prefix $a_1a_2\ldots a_i$ of
$A$ and similarly for $B$.  When $a_i=b_j$, we refer to the pair
$[i,j]$ as a {\em match}; otherwise it is a {\em clash}.

The standard ``naive'' method of computing the length of an LCS is a
``bottom-up'' dynamic programming approach (as in \cite{WagnerF1974})
based on the following recurrence for the length of an LCS of $A_i$
and $B_j$:
 \begin{equation}
 \label{eqn:naiverecur}
 L[i,j] =
    \left\{
    \begin{tabular}{ll}
    0 & if $i=0$ or $j=0$ \\
    $L[i-1,j-1]+1$ & if $i,j>0$ and $a_i=b_j$ \\
    $\max\{L[i-1,j],L[i,j-1]\}$ & otherwise
    \end{tabular}
    \right.
 \end{equation}
 (Sankoff~\cite{Sankoff1972} may be the first to have published this
recurrence, based on the work of Needleman and
Wunsch~\cite{NeedlemanW1970}.)
 In $O(mn)$ time, one may fill an array with all the values of
$L[i,j]$ for $0\leq i\leq m \wedge 0\leq j\leq n$, and the length $L$
of an LCS is read off from $L[m,n]$.  The same time bound also
suffices to produce a single LCS by a ``backtracing'' approach
starting from position $[m,n]$ of the array.  At each stage we just
step from position $[i,j]$ to a position $[i-1,j-1]$, $[i-1,j]$, or
$[i,j-1]$ that is responsible for the setting of $L[i,j]$ as
per~(\ref{eqn:naiverecur}); each match encountered generates a
character of the LCS (in reverse order).

A few other terminologies are useful for discussing some alternative
solution techniques.  Figure~\ref{fig:example} shows the matrix of $L$
values as per~(\ref{eqn:naiverecur}) for a sample pair of
input strings, and we will refer to the value of $L[i,j]$ as the {\em
rank} of $[i,j]$.  It is well known and easy to see that the matches
can be partitioned by rank so as to form {\em contours} as illustrated
by the zig-zag lines in Figure~\ref{fig:example}.  Starting from the
lower left match on a contour, motion along a contour proceeds
monotonically in both dimensions, i.e., the next match is at or above
the level of the previous match and at or to the right of the horizontal
position of the previous match.  Different contours never cross or
touch.  Each contour may be completely specified by the {\em dominant
matches} in the upper left corners of the contours, i.e., those
matches $[\istar,\jstar]$ for which there is no other match $[i',j']$ on the
same contour with
 $i'=\istar \wedge j'<\jstar$ or $j'=\jstar \wedge i'<\istar$.
 For discussion of the algorithm to be presented in
Section~\ref{sec:distinct}, we also introduce the notion of {\em
antidominant matches}, i.e., those matches $[\istar,\jstar]$ for which
there is no other match $[i',j']$ on the same contour with
 $i'=\istar \wedge j'>\jstar$ or $j'=\jstar \wedge i'>\istar$.
 Note also that two matches $[\istar,\jstar]$ and $[i',j']$ with
 $\istar\leq i'$
 can belong to the same common subsequence if and only if
 $\istar<i' \wedge \jstar<j'$.
  Thus, the problem of finding an LCS can be expressed as finding a
longest sequence of matches that is strictly increasing in both
dimensions.

The best known upper bound on the time to find an LCS with general
inputs (i.e., with the time expressed only in terms of $m$ and $n$) is
essentially $O(mn/\lg n)$\cite{MasekP1980} with a finite alphabet or
slightly more with an infinite alphabet~\cite{PatersonD1994}, only a
small improvement over the naive method.  Several other methods have
been proposed to reduce the time under such circumstances as small
alphabet, short LCS, or few dominant matches, e.g.,
\cite{HuntS1977,Hirschberg1977,NakatsuKY1982,HsuD1984,Apostolico1986,ApostolicoG1987,ChinP1990,ApostolicoBG1992,EppsteinGGI1992,Rick2000S}.
For all of these algorithms, however, there are still inputs that
require $\Omega(m^2)$ time or more.\footnote{ An example with many
dominant matches is when one input string contains repeated occurrences
of the pattern {\tt abc} and the other contains repeated occurrences
of the pattern {\tt cba}.}  Thus, the naive method remains a reasonable
approach for finding one LCS, particularly in light of its simplicity.

Relatively little attention has been given to the problem of finding
{\em all} LCS embeddings or {\em all distinct} LCSs.  (In the latter
case, different embeddings of the same character sequence would not be
counted as different LCSs.)  The naive approach to generate all LCS
embeddings~\cite{AhoU1995} would be to extend the backtracing method.
At each step, we would consider three possibilities (and continue
recursively); from position $[i,j]$, we could add a character to the
LCS and move to $[i-1,j-1]$ if $[i,j]$ is a match, and we could move
to $[i-1,j]$ or $[i,j-1]$ if the $L$ value there equals $L[i,j]$
(without adding a character to the LCS and regardless of whether
$[i,j]$ is a match).  One could obviously then remove multiple
embeddings of the same LCS to obtain a list of all distinct LCSs.
This naive approach to generating all LCS embeddings or all distinct
LCSs could, however, be painfully inefficient.  The naive method may
traverse exponentially many paths through the $L$ matrix even when
only one LCS embedding exists.  Furthermore, {\em any} method of
generating distinct LCSs that begins by generating all LCS embeddings
could have a run time exceeding the output size by a factor of
approximately $\frac{3}{\pi n}2.598^n$ as per the maximum number of
different embeddings a single LCS could have in two sequences of
length $n$~\cite{Greenberg2003Btechr}.

Rick~\cite{Rick2000E} gives a method to produce a compact
representation of all LCS embeddings, the {\em LCSs-graph}, from which
all LCS embeddings can be listed in time proportional to the output
size.  He also notes that an extra processing stage can prune the
compact representation to one that gives only distinct LCSs.  The time
complexity of his algorithm for constructing the LCSs-graph $G$ is
$O(\card{\Sigma}n+T+\card{G})$, where $T$ is the time of any algorithm
that determines the dominant matches.  Thus, the run time of his
algorithm is sometimes better than $\Theta(mn)$ but certainly could
require $\Theta(mn)$ for some inputs.  Furthermore, the number of
distinct LCSs could be as large as approximately $1.442^n$ for two
sequences of length $n$\cite{Greenberg2003Btechr}, so actually listing all
distinct LCSs (or all LCS embeddings) may well erase any gain from
constructing the LCSs-graph in less than $\Theta(mn)$ time.
Baeza-Yates\cite{Baezayates1991} provides another construction from
which the LCSs-graph could be produced but with a potentially longer
time of at least $\Theta(\card{\Sigma}n\lg n)$.  $O(mn)$ algorithms for
creating a structure akin to the LCSs-graph have also been proposed by
Gotoh~\cite{Gotoh1990} and Altschul and Erickson~\cite{AltschulE1986}
but with much greater complication than the approach to be presented
here.

This paper shows that a much simpler approach than in prior work can
be used to perform the preprocessing phase in $O(mn)$ time.
Furthermore, the result of this preprocessing phase is a more
versatile structure, the {\em all-prefixes-LCSs-graph}.  From the
all-prefixes-LCSs-graph, we can list, for any prefix $A_i$ of the
first input string and any prefix $B_j$ of the second input string,
all LCSs of $A_i$ and $B_j$ in time proportional to the size of the
output.  The all-prefixes-LCSs-graph can be constructed either for
distinct LCSs or for all LCS embeddings.  The more interesting case of
distinct LCSs is discussed in Section~\ref{sec:distinct}.  The simpler
case of all LCS embeddings is discussed in Section~\ref{sec:all}.

\section{Finding All Distinct Longest Common Subsequences}
 \label{sec:distinct}
 The basic methodology employed here for enabling efficient computation
of all LCSs of any prefixes $A_i$ and $B_j$ of the input sequences is
a variation on the idea of creating a directed acyclic graph in which
every path from the vertex corresponding to $[i,j]$ represents a
different LCS of $A_i$ and $B_j$.  The ``naive backtracing'' approach
mentioned in Section~\ref{sec:intro} can be thought of as directing
edges from $[i,j]$ to some or all of $[i-1,j-1]$, $[i-1,j]$, and
$[i,j-1]$ (according to ranks of the vertices and whether $[i,j]$ is a
match).  Then every path from $[i,j]$ would represent an LCS.  But, as
mentioned before, many paths may be traversed for the same LCS or even
the same LCS embedding; also, throughout the enumeration of paths,
many steps may be taken in which no match is added to the current LCS.

Rick's compact representation of all LCS embeddings in $A$ and $B$,
the LCSs-graph\cite{Rick2000E}, may be defined as follows (though it
is constructed in a much more efficient fashion than this definition
suggests).  Find the transitive closure of the naive graph, remove all
but those vertices that are matches belonging to some LCS, and remove
all edges except those connecting a retained vertex to a retained
vertex of next lower rank.  (Rick actually reverses the direction of
every edge, but declining to do so leaves us in a framework more
analogous to the naive backtracing approach.)  It is easy to see that
the LCSs-graph can be used to list all LCS embeddings in time
proportional to the output size.  Furthermore, Rick notes that a
breadth first search on the graph can be used to eliminate certain
vertices, leaving a representation of only canonical embeddings.  As
noted before, Rick's construction of the LCSs-graph is relatively
complex, and it is predicated upon having first found all dominant
matches.  Furthermore, it is a compact representation only of the LCSs
of $A$ and $B$ rather than of the LCSs for each of the $mn$ pairs of
prefixes $A_i$ and $B_j$.

We show here a simple construction of the all-prefixes-LCSs-graph,
which can be initially thought of as being similar to Rick's
LCSs-graph but without restricting attention to vertices that are
matches belonging to an LCS of $A$ and $B$.  Furthermore, we show how
to prune the edges on the fly so that only the single anticanonical
embedding is represented for each distinct LCS.  Though the number of
edges in the all-prefixes-LCSs-graph may exceed $\Theta(mn)$, we can
still produce essentially an adjacency list representation in $O(mn)$
time, due to a heavy degree of sharing among the adjacency lists of
different vertices.

The precise definition of the all-prefixes-LCSs-graph is as follows.
Every vertex $[i,j]$ has an edge pointing to each match
$[\istar,\jstar]$ of the same rank that is antidominant when
considering the input strings $A_i$ and $B_j$ (i.e., such that
 $\istar\leq i \wedge \jstar\leq j$,
 and there is no other match $[i',j']$ of the same rank with
 $i'=\istar \wedge j\geq j'>\jstar$ or
 $j'=\jstar \wedge i\geq i'>\istar$).
 In Rick's graph, vertices point to vertices of one lower rank, but
the algorithm presented here to generate the all-prefixes-LCSs-graph
is simplified by having vertices point to vertices of equal rank.  It
is still easy to use essentially the same backtracing method to list
the (reversed) LCSs corresponding to any starting point $[i,j]$; we
simply need to augment the explicit edges of the graph with the notion that
whenever we include a match in the LCS, we take a diagonal step
(subtracting one from each matrix coordinate).

Before continuing, we should verify that the backtraces from $[i,j]$
in the all-prefixes-LCSs-graph will actually represent each distinct
LCS once.

\begin{theorem} Considering all paths from $[i,j]$ in the
all-prefixes-LCSs-graph (augmented with diagonal steps from match
nodes) provides a one-to-one correspondence with distinct LCSs of
$A_i$ and $B_j$.
\end{theorem}

\begin{proof}
 This is easiest to see by recalling that finding all LCS embeddings
(in reverse order) corresponds to finding all longest sequences of
matches (in the submatrix defined by $A_i$ and $B_j$) that are
strictly decreasing in both dimensions.

The restricted use of matches that we incorporated into the
all-prefixes-LCSs-graph will not make us lose any LCSs, by the
following reasoning.  If we consider a backtrace in which we go from
position $[\itilde,\jtilde]$ to a match $[\istar,\jstar]$ such that
there exists another match $[i',j']$ of the same rank with
 $i'=\istar \wedge j\geq j'>\jstar$
 or
 $j'=\jstar \wedge i\geq i'>\istar$,
 then we can just replace $[\istar,\jstar]$ with $[i',j']$ and get the
same LCS.  ($[\istar,\jstar]$ and $[i',j']$ must match on the same
character, since they are in the same row or column.)

We also will not duplicate any LCSs, by the following reasoning.  To
get the same LCS twice, there would need to be a position
$[\itilde,\jtilde]$ from which two edges in the all-prefixes-LCS-graph
proceed to two matches on the same character, say $[\istar,\jstar]$
and $[\ihat,\jhat]$ with $\ihat\geq\istar$ and $\jstar>\jhat$.  Then
$[\ihat,\jstar]$ would be a match on the same character and of the
same rank, implying that $[\istar,\jstar]$ or $[\ihat,\jhat]$ is not
antidominant for $A_i$ and $B_j$.
 \end{proof}

\begin{figure}
 \centering
  \setlength{\unitlength}{.4em}
 {\small
 \renewcommand{\thicklines}{\linethickness{.3ex}}
 \renewcommand{\thinlines}{\linethickness{.05ex}}
 \thicklines
 \newcommand{\node}{\rect(12,4)}
 \newcommand{\redge}{\put(6,0){\vector(1,0){4}}}
 \newcommand{\dedge}{\put(0,-2){\vector(0,-1){4}}}
 \newcommand{\uedge}{\put(0,2){\vector(0,1){4}}}
 \newcommand{\dr}{\thinlines\put(5,-2){\vector(1,-1){4}}\thicklines}
 \newcommand{\ur}{\thinlines\put(5,2){\vector(1,1){4}}\thicklines}
 \begin{picture}(109,20)(-9,-10)
 \put(3.5,0){\makebox(0,0)[r]{$\var{head}[i,j]$}}
 \put(-2,0){\redge}
 \put(14,0){\node\redge}
 \put(30,0){\node\redge}
 \put(46,0){\node\redge}
 \put(62,0){\node\redge}
 \put(78,0){\node\redge}
 \put(94,0){\node}
 \put(23.5,8){\makebox(0,0)[r]{$\var{head}[i',j']$}}
 \put(18,8){\redge}
 \put(34,8){\node}
 \put(40,8){\dedge}
 \put(55.5,-6){\makebox(0,0)[tr]{$\var{pretail}[i,j]$}}
 \put(56,-8){\uedge}
 \put(67.5,-8){\makebox(0,0)[r]{$\var{tail}[i,j]$}}
 \put(62,-8){\redge}
 \put(63,0){\dr}
 \put(78,-8){\node}
 \put(71.5,6){\makebox(0,0)[br]{$\var{pretail}[i',j']$}}
 \put(72,8){\dedge}
 \put(83.5,8){\makebox(0,0)[r]{$\var{tail}[i',j']$}}
 \put(78,8){\redge}
 \put(79,0){\ur}
 \put(94,8){\node}
 \end{picture}
 }
 \caption{Adjacency lists of different nodes in the
all-prefixes-LCSs-graph might share portions of a contour list.  In
this example, the contour list could be specified by the central line
of ``next node'' pointers, while the adjacency lists of $[i,j]$ and
$[i',j']$ are largely just excerpts.  The two thin diagonal lines are
not actual pointers present in the data structure; rather they show
the implicit relationships that $\pretail[i,j]$ is followed by
$tail[i,j]$ in the adjacency list for $[i,j]$, and similarly for
$[i',j']$.}
 \label{fig:linkedlists}
 \end{figure}
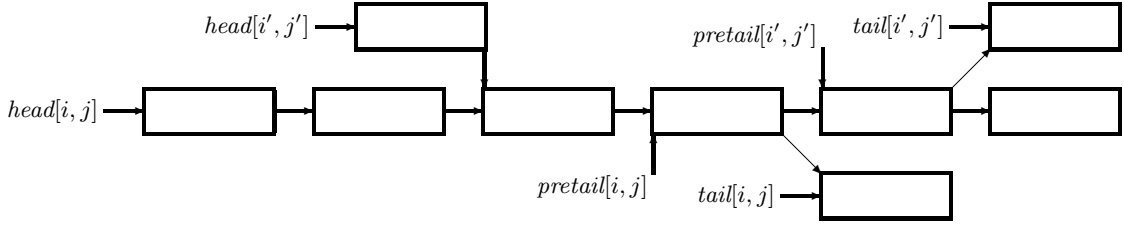

Now, each adjacency list in the all-prefixes-LCSs-graph can be thought
of as being based upon a linked list of the antidominant matches along
one of the contours; let us call such a linked list a {\em contour
list}.  That is, the antidominant matches for $A$ and $B$ are nearly
sufficient to characterize all distinct LCSs for each $A_i$ and $B_j$
(whereas the {\em dominant} matches do not suffice, as is illustrated
by Rick~\cite{Rick2000E}).  We need only add to the adjacency list for
$[i,j]$ at most two matches that are not antidominant for $A$ and $B$,
by considering the extreme lower left and upper right matches
$[\istar,\jstar]$ on the relevant contour satisfying
 $\istar\leq i \wedge \jstar\leq j$.  Thus, each adjacency list in the
all-prefixes-LCSs-graph is a portion of a contour list, with the
possible addition of a different head and/or tail node.  An adjacency
list that has a separate head node and then jumps into the midst of a
contour list can easily share all but a constant amount of its storage
with the contour list; see Figure~\ref{fig:linkedlists}.  When
adjacency lists digress to incorporate a separate tail node, however,
we require a small digression from the standard linked list
representation.  It suffices to maintain, for each adjacency list, a
pointer to the second to last node $\pretail[i,j]$ as well as a
pointer to the last node $\tail[i,j]$, where $\tail[i,j]$ may not
actually be the target of an ordinary linked list next node pointer
from $\pretail[i,j]$.  (See Figure~\ref{fig:linkedlists}.)

With the above representation in mind, we can adapt the naive method
of calculating LCS length based on~(\ref{eqn:naiverecur})
to also set up the appropriate $\head[i,j]$, $\pretail[i,j]$, and
$\tail[i,j]$ pointers based on the information already computed at
positions $[i-1,j-1]$, $[i-1,j]$, and $[i,j-1]$.  We always use
$\tail[i,j]$ to point to the last node on the adjacency list at
position $[i,j]$ or assign \NULL\ for an empty adjacency list.  Any
{\em other} nodes in the adjacency list appear in an ordinary linked
list beginning at $\head[i,j]$ and terminating at $\pretail[i,j]$,
with $\head[i,j]$ being \NULL\ if there are no such nodes.

The algorithm to construct the all-prefixes-LCSs-graph in $O(mn)$ time
is given in Figure~\ref{fig:distinct-alg}.  We proceed through
positions $[i,j]$ in a row-by-row fashion, with trivial handling for
matches in line~\ref{line:match} and handling of a clash in
Lines~\ref{line:clash-begin} to~\ref{line:clash-end}.

\begin{figure}
 \centering
 \input{fig-distinct-alg}
 \caption{The algorithm to create the all-prefixes-LCSs-graph
representing all distinct LCSs.}
 \label{fig:distinct-alg}
 \end{figure}

When $[i,j]$ is a match; the corresponding vertex in the
all-prefixes-LCSs-graph just points to itself; any other match of the
same rank must have a higher row or column index.  Consistent with the
adjacency list approach described above, we use $\tail[i,j]$ for the
pointer to self and leave $\head[i,j]$ \NULL\ to indicate there is
nothing else in the adjacency list of $[i,j]$.

When $[i,j]$ is a clash, backtraces in the naive graph need only
proceed through whichever of $[i-1,j]$ and $[i,j-1]$ are of the same
rank as $[i,j]$.  To avoid following duplicate paths that uncover the
same embedding or uncovering multiple embeddings of a single LCS, we
perform an appropriate merging of information computed at $[i-1,j]$
and $[i,j-1]$.  The adjacency lists are always maintained so that the
nodes are in order along a contour from lower left to upper right, and
the adjacency lists at $[i-1,j]$ and $[i,j-1]$ are identical (if
$[i-1,j]$ and $[i,j-1]$ are of the same rank) except for possibly a
different head and/or tail.  Once we conditionally set the tail of the
adjacency list for position $[i,j]$ from the information at $[i-1,j]$
in line~\ref{line:up}, we need only look for additional information at
position $[i,j-1]$ in lines~\ref{line:left-begin}
to~\ref{line:left-end}.

If a null $\tail[i,j]$ was obtained by looking at position $[i-1,j]$,
then the adjacency list for $[i,j]$ contains at most one node; in that
case $\tail[i,j]$ is copied from position $[i,j-1]$, and our work is
done.  Otherwise, we can tentatively form the adjacency list of
$[i,j]$ by just following the adjacency list of $[i,j-1]$ with
$\tail[i,j]$.  But we need
lines~\ref{line:pare-shadowed-begin}--~\ref{line:pare-shadowed-end} to
ensure that each LCS is only represented once; these lines strip from
this tentative adjacency list any matches that are duplicative or are
not antidominant.  If the condition of
line~\ref{line:pare-shadowed-begin} is satisfied, then $\tail[i,j-1]$
must be stripped out; in any case, $\pretail[i,j]$ can be given the
correct value for the case that the adjacency list of $[i,j]$ contains
more than just $\tail[i,j]$.  If the condition of
line~\ref{line:pare-shadowed-end} is satisfied, the tentative value of
$\tail[i,j]$ must be stripped out; since it was not stripped out when
considering position $[i-1,j]$, $\tail[i,j]$ must be in row $i$, and
only a single node remains in the adjacency list for $[i,j]$, so that
our work is again done.  Otherwise, we set the next linked list node
of $\pretail[i,j]$ to $\tail[i,j]$ in line~\ref{line:link}, and we set
$\head[i,j]$ in lines~\ref{line:head-set-begin}
to~\ref{line:head-set-end}.  (The explicit link from $\pretail[i,j]$
to $\tail[i,j]$ will not be used when processing the adjacency list of
$[i,j]$; it may be overwritten when a higher value of $j$ is
considered, or it may then become relevant due to $\pretail$ advancing
along the relevant contour list.  As written, the algorithm does not
guarantee full construction of all contour lists, but it constructs
the portions needed in the adjacency lists of the
all-prefixes-LCSs-graph.)

\section{Finding All Longest Common Subsequence Embeddings}
 \label{sec:all}
 If we wish to find all LCS embeddings, without duplicating the same
embedding, but including multiple embeddings of the same LCS, we can
follow an approach similar to that of Section~\ref{sec:distinct} but
with some simplification.  We can now have each vertex $[i,j]$ in the
all-prefixes-LCS-graph point to {\em all} matches $[\istar,\jstar]$ of
the same rank with
 $\istar\leq i \wedge \jstar\leq j$.
  It is no longer necessary to incorporate $\pretail$ information to
specify the adjacency lists; rather, each adjacency list is simply an
excerpt of a contour list, and each can be specified with a $\head$
and $\tail$ pointer.

It is now relatively easy to see that this version of the
all-prefixes-LCSs-graph can be constructed in $O(mn)$ time as per the
algorithm in Figure~\ref{fig:embeddings-alg}.  We now use
line~\ref{line:match2} to set self-pointers for a match.  Then,
regardless of whether $[i,j]$ is a match, we merge the information
there with information at whichever of $[i-1,j]$ and $[i,j-1]$ is of
the same rank.  In
lines~\ref{line:up-body-begin}--\ref{line:up-body-end}, the adjacency
list for $[i,j]$ becomes that of $[i-1,j]$ with $[i,j]$ possibly added
on front.  In
lines~\ref{line:left-body-begin}--\ref{line:left-body-end}, the
adjacency list for $[i,j-1]$ (known to be nonempty due to the test in
line~\ref{line:left-test}) is merged onto the front of the adjacency
list for $[i,j]$ (taking into account possible overlap).

\begin{figure}
 \centering
 \input{fig-embeddings-alg}
 \caption{The algorithm to create the all-prefixes-LCSs-graph
representing all LCS embeddings.}
 \label{fig:embeddings-alg}
 \end{figure}

\section{Conclusion}
 \label{sec:conc}
 Simple $O(mn)$ algorithms have been presented to produce the
all-prefixes-LCS-graph in either the context of representing all
distinct LCSs or of representing all LCS embeddings.  Once this graph
is constructed, we can list all the LCSs (or LCS embeddings) of
prefixes $A_i$ and $B_j$ of the two input strings in time proportional
to the output size.  A {\sc C} language implementation closely
following the presentation given in this paper can be found at
\url{http://www.cs.luc.edu/~rig/lcs}.  Also included is an
implementation of the ``naive backtracing'' method followed by removal of
duplicate LCSs (or duplicate embeddings).  Results for many examples
are included, with the same final list of LCSs (or LCS embeddings)
resulting from the algorithms in this paper or the naive method
followed by removal of duplicates.

\bibliographystyle{hplain}
\bibliography{sources}

\end{document}